\documentclass{elsartNoFoot}
\usepackage{amssymb}
\usepackage[pdftex]{color}

\definecolor{red  }{rgb}{0.8,0.0,0.0}
\definecolor{green}{rgb}{0.0,0.8,0.0}

\newcommand{\1}{\color{red}}
\newcommand{\2}{\color{green}}

\renewcommand{\leq}{\leqslant}		
\renewcommand{\geq}{\geqslant}		

\newcommand{\set}[1]{\{#1\}}		

\newcommand{\X}[1]{x_{#1}}		
\renewcommand{\S}[1]{s_{#1}}		

\newcommand{\+}[3]{%
        {%
        \renewcommand{\i}{{#1}}%
        {#3}%
        ,\ldots,%
        \renewcommand{\i}{{#2}}%
        {#3}%
        }%
}

\newcommand{\ti}{{\tt i}}		
\newcommand{\tk}{{\tt k}}		

\renewcommand{\exp}[1]{\gcd(N,#1)}	

\newcommand{\rf}[2]{\stackrel{(\ref{#1})}{#2}}

\newcommand{\SUM}[1]{\Sigma_{#1}}

\begin{document}


\begin{frontmatter}

\title{Maintaining partial sums in logarithmic time}
\author{Jochen Burghardt}
\address{Berlin}

\begin{abstract}
        We present a data structure that allows to maintain in
        logarithmic time all partial sums of elements of a linear array
        during incremental changes of element's values. 
\end{abstract}

\end{frontmatter}

\section{Motivation}
\label{Motivation}

Assume you have a linear array $\+0{N-1}{\X{\i}}$ of numbers
which are frequently updated, and you need to maintain all partial
sums $\sum_{i=j}^k \X{i}$, where $0 \leq j \leq k < N$.
We present a data structure that allows to access each $\X{k}$ and to
compute any partial sum in time ${\mathcal O}(\log(N))$.

As an application, think of the $\X{k}$
as integer numbers indicating the probabilities of certain
events; 
by chosing a uniformly distributed random number $r$ in the range
$0 \leq r < \sum_{i=0}^{N-1} \X{i}$
and selecting the unique 
$k \in \set{\+0N\i}$
with
$\sum_{i=k}^{N-1} \X{i} \leq r < \sum_{i=k-1}^{N-1} \X{i}$,
event $k$ is selected with probability
$\frac{\X{k}}{\sum_{i=0}^{N-1} \X{i}}$.

If the probability distribution of events changes frequently,
the partial sums need to be recomputed every time,
which takes time ${\mathcal O}(N)$ using the naive algorithm.

\begin{figure}
\begin{center}
\begin{picture}(8.3,4.5)
\put(0.242,1.210){\makebox(0.000,0.000){$i$}}
\put(0.847,1.210){\makebox(0.000,0.000){$\scriptstyle 0$}}
\put(1.331,1.210){\makebox(0.000,0.000){$\scriptstyle 1$}}
\put(1.815,1.210){\makebox(0.000,0.000){$\scriptstyle 2$}}
\put(2.299,1.210){\makebox(0.000,0.000){$\scriptstyle 3$}}
\put(2.783,1.210){\makebox(0.000,0.000){$\scriptstyle 4$}}
\put(3.267,1.210){\makebox(0.000,0.000){$\scriptstyle 5$}}
\put(3.751,1.210){\makebox(0.000,0.000){$\scriptstyle 6$}}
\put(4.235,1.210){\makebox(0.000,0.000){$\scriptstyle 7$}}
\put(4.719,1.210){\makebox(0.000,0.000){$\scriptstyle 8$}}
\put(5.203,1.210){\makebox(0.000,0.000){$\scriptstyle 9$}}
\put(5.687,1.210){\makebox(0.000,0.000){$\scriptstyle 10$}}
\put(6.171,1.210){\makebox(0.000,0.000){$\scriptstyle 11$}}
\put(6.655,1.210){\makebox(0.000,0.000){$\scriptstyle 12$}}
\put(7.139,1.210){\makebox(0.000,0.000){$\scriptstyle 13$}}
\put(7.623,1.210){\makebox(0.000,0.000){$\scriptstyle 14$}}
\put(8.107,1.210){\makebox(0.000,0.000){$\scriptstyle 15$}}
\put(0.242,0.726){\makebox(0.000,0.000){$\S{i}$}}
\put(0.847,0.726){\makebox(0.000,0.000){$99$}}
\put(1.331,0.726){\makebox(0.000,0.000){$8$}}
\put(1.815,0.726){\makebox(0.000,0.000){$9$}}
\put(2.299,0.726){\makebox(0.000,0.000){$3$}}
\put(2.783,0.726){\makebox(0.000,0.000){$17$}}
\put(3.267,0.726){\makebox(0.000,0.000){$1$}}
\put(3.751,0.726){\makebox(0.000,0.000){$8$}}
\put(4.235,0.726){\makebox(0.000,0.000){$3$}}
\put(4.719,0.726){\makebox(0.000,0.000){$51$}}
\put(5.203,0.726){\makebox(0.000,0.000){$7$}}
\put(5.687,0.726){\makebox(0.000,0.000){$7$}}
\put(6.171,0.726){\makebox(0.000,0.000){$4$}}
\put(6.655,0.726){\makebox(0.000,0.000){$17$}}
\put(7.139,0.726){\makebox(0.000,0.000){$2$}}
\put(7.623,0.726){\makebox(0.000,0.000){$9$}}
\put(8.107,0.726){\makebox(0.000,0.000){$5$}}
\put(0.242,0.242){\makebox(0.000,0.000){$\X{i}$}}
\put(0.847,0.242){\makebox(0.000,0.000){$\scriptstyle 14$}}
\put(1.331,0.242){\makebox(0.000,0.000){$\scriptstyle 8$}}
\put(1.815,0.242){\makebox(0.000,0.000){$\scriptstyle 6$}}
\put(2.299,0.242){\makebox(0.000,0.000){$\scriptstyle 3$}}
\put(2.783,0.242){\makebox(0.000,0.000){$\scriptstyle 8$}}
\put(3.267,0.242){\makebox(0.000,0.000){$\scriptstyle 1$}}
\put(3.751,0.242){\makebox(0.000,0.000){$\scriptstyle 5$}}
\put(4.235,0.242){\makebox(0.000,0.000){$\scriptstyle 3$}}
\put(4.719,0.242){\makebox(0.000,0.000){$\scriptstyle 20$}}
\put(5.203,0.242){\makebox(0.000,0.000){$\scriptstyle 7$}}
\put(5.687,0.242){\makebox(0.000,0.000){$\scriptstyle 3$}}
\put(6.171,0.242){\makebox(0.000,0.000){$\scriptstyle 4$}}
\put(6.655,0.242){\makebox(0.000,0.000){$\scriptstyle 6$}}
\put(7.139,0.242){\makebox(0.000,0.000){$\scriptstyle 2$}}
\put(7.623,0.242){\makebox(0.000,0.000){$\scriptstyle 4$}}
\put(8.107,0.242){\makebox(0.000,0.000){$\scriptstyle 5$}}
\put(8.107,1.573){\circle*{0.182}}
\put(7.139,1.573){\circle*{0.182}}
\put(6.171,1.573){\circle*{0.182}}
\put(5.203,1.573){\circle*{0.182}}
\put(4.235,1.573){\circle*{0.182}}
\put(3.267,1.573){\circle*{0.182}}
\put(2.299,1.573){\circle*{0.182}}
\put(1.331,1.573){\circle*{0.182}}
\put(7.623,2.178){\circle*{0.182}}
\put(5.687,2.178){\circle*{0.182}}
\put(3.751,2.178){\circle*{0.182}}
\put(1.815,2.178){\circle*{0.182}}
\put(6.655,2.783){\circle*{0.182}}
\put(2.783,2.783){\circle*{0.182}}
\put(4.719,3.388){\circle*{0.182}}
\put(0.847,4.174){\circle*{0.182}}
\put(8.107,1.573){\line(-4,5){0.484}}
\put(7.139,1.573){\line(-4,5){0.484}}
\put(6.171,1.573){\line(-4,5){0.484}}
\put(5.203,1.573){\line(-4,5){0.484}}
\put(4.235,1.573){\line(-4,5){0.484}}
\put(3.267,1.573){\line(-4,5){0.484}}
\put(2.299,1.573){\line(-4,5){0.484}}
\put(1.331,1.573){\line(-4,5){0.484}}
\put(7.623,1.573){\line(0,1){0.605}}
\put(6.655,1.573){\line(0,1){1.210}}
\put(5.687,1.573){\line(0,1){0.605}}
\put(4.719,1.573){\line(0,1){1.815}}
\put(3.751,1.573){\line(0,1){0.605}}
\put(2.783,1.573){\line(0,1){1.210}}
\put(1.815,1.573){\line(0,1){0.605}}
\put(0.847,1.573){\line(0,1){2.662}}
\put(7.623,2.178){\line(-5,3){0.967}}
\put(5.687,2.178){\line(-5,3){0.967}}
\put(3.751,2.178){\line(-5,3){0.967}}
\put(1.815,2.178){\line(-5,3){0.967}}
\put(6.655,2.783){\line(-3,1){1.936}}
\put(2.783,2.783){\line(-3,1){1.936}}
\put(4.719,3.388){\line(-5,1){3.920}}
\end{picture}
\end{center}
\caption{Data structure}
\label{tree}
\end{figure}
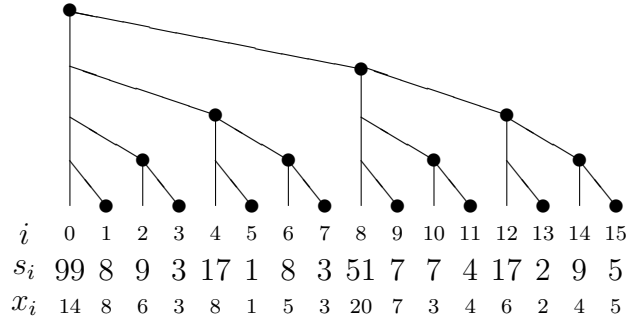

\section{Data structure and access algorithms}
\label{Data structure and access algorithms}

Our solution is to store a mix of individual values $\X{i}$ and partial
sums in the array, thus realizing a binary tree where each node
represents the sum of all leafs below it.
Figure~\ref{tree}
sketches an example for $N=16$, the partial sums
corresponding to the nodes
indicated by solid circles are stored as $\S{i}$.

In some respect, this idea is similar to that of heap sort
\cite[Sect.~3.4]{Aho.Hopcroft.Ullman.1974}, which also uses a mix of
representations (sorted along a path and unsorted within a level) to
combine the advantages of both. Our data structure combines the
advantages of storing single values (easily updatable) and sums (no need
to recompute them).

Formally, let $N$ be a power of 2;
let an array $\+0{N-1}{\X{\i}}$ of size $N$ be given.
Instead of this original array, 
we maintain the array $\+0{N-1}{\S{\i}}$,
where

\begin{equation}
\S{k} := \sum_{i=0}^{\exp{k}-1} \X{k+i}		\label{dataInv}
\end{equation}

Here, $\exp{k}$ is 
the greatest common divisor of $N$ and $k$, i.e.,
the largest power of 2 dividing $k$.
It corresponds to the least 1 bit in the 2--complement
representation of $k$, which can be computed as bitwise {\tt and}
of $N+k$ and $N-k$.

\begin{figure}
\begin{center}
\begin{verbatim}
int s[M];

#define S(i)              (i<M ? s[i] : 0)

#define gcdN(k)           ((N+k) & (N-k))

int sumN(int k) {
    int i, sm = 0;
    for (i=k; i<M; i+=gcdN(i))
        sm += s[i];
    return sm;
}

#define sum(j,k)          (sumN(j) - sumN(k+1))

int get(int k) {
    int i, x = s[k];
    for (i=1; i<gcdN(k) && k+i<M; i*=2)
        x -= s[k+i];
    return x;
}

void inc(int k,x) {
    int i;
    for (i=k; i>=0; i-=gcdN(i))
        s[i] += x;
}

#define set(k,x)          inc(k,x-get(k))

int find(int x) {
    int i, k = 0, pv = s[N/2];
    for (i=N/2; i>0; i/=2)
        if (x < pv) {
            pv += S(k+i*3/2) - s[k+i];
            k  += i;
        } else {
            pv += S(k+i/2);
        }
    return k;
}
\end{verbatim}
\caption{Algorithms}
\label{Algorithms}
\end{center}
\end{figure}

\begin{figure}%
\begin{center}%
{\small\tt%
\begin{tabular}[t]{@{}*{12}{|r}|@{}}%
\multicolumn{6}{@{}l@{}}{\rule{0cm}{0.5cm}{\tt sumN(3)}}     
	& \multicolumn{1}{l}{} &
	\multicolumn{5}{@{}l@{}}{{\tt get(12)}}       \\
\cline{1-6}
	\cline{8-12}
i & & 3 & 4 & 8 & 16    &&
	i & & 1 & 2 & 4 \\
\cline{1-6}
	\cline{8-12}
sm & 0 & 3 & 20 & 71 &	&&
	x & 17 & 15 & 6 &   \\
\cline{1-6}
	\cline{8-12}
\multicolumn{4}{@{}l@{}}{\rule{0cm}{0.5cm}{\tt inc(12,...)}}
	& \multicolumn{1}{l}{} &
	\multicolumn{7}{@{}l@{}}{\rule{0cm}{0.5cm}{\tt find(69)}}     \\
\cline{1-4}
	\cline{6-12}
i & 12 & ~8 & 0        &&
	i   &    & 8  & 4  & 2  & 1  & ~0	\\
\cline{1-4}
	\cline{6-12}
\multicolumn{4}{@{}l@{}}{\rule{0cm}{0.5cm}{\tt inc(3,...)}}
	&&
	pv  & 51 & 68 & 77 & 71 & 71 &	\\
\cline{1-4}
	\cline{6-12}
i & 3 & 2 & 0	&&
	k   & 0  &    &    & 2  & 3  &	\\
\cline{1-4}
	\cline{6-12}
\end{tabular}%
}%
\end{center}%
\caption{Sample runs}%
\label{Sample runs}%
\end{figure}


The following algorithms, given in C code
in Fig.~\ref{Algorithms}, maintain our data structure.

\begin{itemize}
\item \verb|int  sumN(int k) |
	returns $\sum_{i=\tk}^{N-1} \X{i}$;
\item \verb|int  sum(int j,k)|
	returns $\sum_{i={\tt j}}^\tk \X{i}$;
\item \verb|int  get(int k)  |
	retrieves $\X{\tk}$;
\item \verb|void inc(int k,x)|
	adds {\tt x} to $\X{\tk}$;
\item \verb|void set(int k,x)|
	assigns {\tt x} to $\X{\tk}$; and
\item \verb|int  find(int x) |
	returns some $k$ such that
	$\sum_{i=k+1}^{N-1}\X{i} \leq {\tt x} < \sum_{i=k}^{N-1}\X{i}$,
	\\
	provided 
	$0 \leq {\tt x} < \sum_{i=0}^{N-1}\X{i}$;
	~
	$k$ is unique if no $\X{i}$ is negative.
\end{itemize}

Figure \ref{Sample runs} shows some sample runs on the data in
Fig.~\ref{tree}.

In order to deal with arrays whose size is not a
power of 2, assume $\S{k} = 0$ for all $k \geq M$, where
$N/2 < M \leq N$. At two places it is neccessary to test
the index boundary explicitely, using the function {\tt int S(int i)}.

The algorithms can immediately be generalized to deal with arbitrary
(non--abelian) group elements instead of integers; if {\tt find} is to
be used, ordered groups are neccessary.

\section{Complexity}
\label{Complexity}

All algorithms take ${\mathcal O}(\log N)$ time due to the implicit tree
structure.
For {\tt sumN} and {\tt inc}, note that the value of $\exp{\ti}$
grows in every loop cycle, since

\begin{equation}
	\exp{i \pm \exp{i}} \geq 2 \cdot \exp{i} \; .
	\label{expIncr}
\end{equation}

In the following sections \ref{Correctness of get} to
\ref{Correctness of find}, we give correctness proofs of the main
algorithms in the Hoare calculus \cite{Hoare.1969}.

\section{Correctness of {\tt get}}
\label{Correctness of get}

To see the correctness of {\tt get}, show

\begin{equation}
	\sum_{i=1}^{2^a-1} f(i)
	= \sum_{i=0}^{a-1} \sum_{j=0}^{2^i-1} f(2^i+j)	
	\label{reord}
\end{equation}

by induction on $a$;
note that commutativity of $+$ is not required for the proof.

If $\exp{k} = 2^a$, we have $\exp{k+2^i} = 2^i$ for $0 \leq i < a$,
and therefor 

\begin{equation}
\S{k} 
\rf{dataInv}= \X{k} + \sum_{i=1}^{2^a-1} \X{k+i}
\rf{reord}= \X{k} + \sum_{i=0}^{a-1} \sum_{j=0}^{2^i-1} \X{k+2^i+j}
\rf{dataInv}= \X{k} + \sum_{i=0}^{a-1} \S{k + 2^i} \; .
\label{recurs}
\end{equation}

We define the abbreviation
$\SUM{\tk,b} 
:= \S{\tk+b} + \S{\tk+2\cdot b} + \S{\tk+4\cdot b} 
+ \ldots + \S{\tk+\exp{\tk}/2}$ .

By equation (\ref{recurs}), we obtain 
$\S{\tk} = \X{\tk} + \SUM{\tk,1}$, 
justifying the step in lines 4.--5.

We have $\SUM{\tk,b} = 0$ if $b \geq \exp{\tk}$ or
$\tk+b \geq M$; this justifies lines 13.--14.

We can now apply the Hoare calculus to the code of
\verb|int get(int k)|:

\begin{enumerate}
\setlength{\itemsep}{-0.0cm}
\renewcommand{\labelenumi}{\arabic{enumi}.}
 \item{\1\verb|int get(int k) {					|
}\item{\1\verb|    int i, x;					|
}\item{\1\verb|    x = s[k];					|
}\item{\1\verb|    i = 1;					|
}\item{\2\verb|    |%
	${\tt x} = \X{\tk} + \SUM{\tk,\ti}
	\wedge \ti \leq \exp{\tk}$
}\item{\1\verb|    while (i < gcdN(k) && k+i < M) {		|
}\item{\2\verb|        |%
	${\tt x} = \X{\tk} + \SUM{\tk,\ti}
	\wedge \ti < \exp{\tk}$
}\item{\1\verb|        x = x - s[k+i];				|
}\item{\2\verb|        |%
	${\tt x} = \X{\tk} + \SUM{\tk,2\cdot\ti}
	\wedge \ti < \exp{\tk}$
}\item{\1\verb|        i = i * 2;				|
}\item{\2\verb|        |%
	${\tt x} = \X{\tk} + \SUM{\tk,\ti}
	\wedge \ti \leq \exp{\tk}$
}\item{\1\verb|    }						|
}\item{\2\verb|    |%
	${\tt x} = \X{\tk} + \SUM{\tk,\ti}
	\wedge (\ti = \exp{\tk} \vee \tk+\ti \geq M)$
}\item{\1\verb|    return x;					|
}\item{\1\verb|}						|
}
\end{enumerate}

\section{Correctness of {\tt inc}}
\label{Correctness of inc}

Next, we show that {\tt inc} makes sufficiently many updates.
By (\ref{dataInv}),
$\S{i}$ depends on $\X{k}$, 
iff
$i \leq k < i+\exp{i}$.

Hence, if $\S{i}$ depends on $\X{k}$,
then so does $\S{i-\exp{i}}$, since

\begin{eqnarray*}
	\begin{array}{@{}rcccll@{}}
	i - \exp{i} & \leq & i & \leq & k
	& \mbox{and, by (\ref{expIncr}),}	\\
	(i - \exp{i}) + \exp{i - \exp{i}}
	& \geq & 
	i + \exp{i} & > & k & .	\\
	\end{array}
\end{eqnarray*}

But no $\S{i'}$ for $i - \exp{i} < i' < i$ depends on $\X{k}$:
\begin{quote}
Let $i = 2^a \cdot b$ and $i' = 2^{a'} \cdot b'$ 
for odd numbers $b, b'$.
\\
Then $a' < a$ since $i - \exp{i} = 2^a \cdot (b-1)$.
\\
And $2^{a'} \cdot b' = i' < i = 2^{a-a'} \cdot 2^{a'} \cdot b$
implies $b'+1 \leq 2^{a-a'}\cdot b$.
\\
Hence,
$
	i' + \exp{i'} 
	= 2^{a'} \cdot (b'+1) 
	\leq 2^{a'} \cdot 2^{a-a'}\cdot b
	= i
	\leq k \; .
$
\end{quote}

\section{Correctness of {\tt sumN}}
\label{Correctness of sumN}

The loop in {\tt sumN} satisfies the invariant
${\tt sm} = \sum_{j=\tk}^{\ti-1} \X{j}$,
since

\begin{displaymath}
{\tt sm} + \S{i}
\rf{dataInv}= \left( \sum_{j=\tk}^{\ti-1} \X{j} \right)
+ \left( \sum_{j=0}^{\exp{\ti}-1} \X{\ti+j} \right)
= \sum_{j=\tk}^{\ti+\exp{\ti}-1} \X{j} \; .
\end{displaymath}

This justifies the step in lines 7.--9.
For lines 13.--14.\ note that $\X{j} = 0$ for $j \geq M$.

\begin{enumerate}
\setlength{\itemsep}{-0.0cm}
\renewcommand{\labelenumi}{\arabic{enumi}.}
 \item{\1\verb|int sumN(int k) {				|
}\item{\1\verb|    int i, sm;					|
}\item{\1\verb|    sm = 0;					|
}\item{\1\verb|    i  = k;					|
}\item{\2\verb|    |%
	${\tt sm} = \sum_{j=\tk}^{\ti-1} \X{j}$
}\item{\1\verb|    while (i < M) {				|
}\item{\2\verb|        |%
	${\tt sm} = \sum_{j=\tk}^{\ti-1} \X{j}$
}\item{\1\verb|        sm = sm + s[i];				|
}\item{\2\verb|        |%
	${\tt sm} = \sum_{j=\tk}^{\ti+\exp{\ti}-1} \X{j}$
}\item{\1\verb|        i  = i  + gcdN(i);			|
}\item{\2\verb|        |%
	${\tt sm} = \sum_{j=\tk}^{\ti-1} \X{j}$
}\item{\1\verb|    }						|
}\item{\2\verb|    |%
	${\tt sm} = \sum_{j=\tk}^{\ti-1} \X{j}
	\wedge \ti \geq M$
}\item{\1\verb|    return sm;					|
}\item{\1\verb|}						|
}
\end{enumerate}

\section{Correctness of {\tt find}}
\label{Correctness of find}

The loop in {\tt find} satisfies the invariant

\begin{eqnarray}
	\begin{array}{lll}
	\sum_{j=\tk+2 \cdot \ti}^{N-1} \X{j}
		\leq {\tt x}
		< \sum_{j=\tk}^{N-1} \X{j}
	& \mbox{ ~ and ~ } &
	\left( \ti \geq 2 \Rightarrow {\tt pv}
		= \sum_{j=\tk+\ti}^{N-1} \X{j} \right)	\\
	& \mbox{ ~ and ~ } & \exp{k} \geq \exp{i} = i \; .
	\end{array}
\label{findInv}
\end{eqnarray}

To show this, note that for $i \geq 2$, we have

\begin{displaymath}
\S{\tk+\ti} 
\rf{dataInv}= \sum_{j=0}^{\exp{\tk + \ti}-1} \X{\tk + \ti + j}
\rf{findInv}= \sum_{j=0}^{\ti - 1} \X{\tk + \ti + j}
= \sum_{j=\tk+\ti}^{\tk + 2 \cdot \ti - 1} \X{j} \; ,
\end{displaymath}

and similarly

\begin{displaymath}
	\S{\tk+\ti / 2} = \sum_{j=\tk+\ti / 2}^{\tk + \ti - 1} \X{j} 
		\; \mbox{ ~ and ~ } \;
	\S{\tk+\ti \cdot 3 / 2} 
		= \sum_{j=\tk+\ti \cdot 3 / 2}^{\tk + 2 \cdot \ti - 1}
		\X{j}
		\; ,
\end{displaymath}

hence, we get

\begin{equation}
	{\tt pv} + \S{\tk+\ti \cdot 3 / 2} - \S{\tk+\ti}
	= \sum_{j=\tk+\ti \cdot 3 / 2}^{N-1} \X{j}
	\; \mbox{ ~ and ~ } \;
	{\tt pv} + \S{\tk+\ti / 2}
	= \sum_{j=\tk+\ti / 2}^{N-1} \X{j} \; ,
	\label{pvHoare}
\end{equation}

in case of ${\tt x} < {\tt pv}$ and ${\tt x} \geq {\tt pv}$,
respectively.

We transform the program to make the Hoare verification rules
applicable and unfold the last loop cycle ($\ti = 1$) to avoid
confusing case distinctions.
We omit the computation of the pivot element
{\tt pv} in the last cycle, since its value
isn't used any more.

We define the abbreviations
~
$\SUM{a} := \sum_{j=a}^{N-1} \X{j}$
~
and
~
$p(a,b) :\Leftrightarrow \exp{a} \geq \exp{b} = b$

Observe that 
$\ti \geq 2 \wedge p(\tk,\ti)$
implies both $p(\tk+\ti,\ti)$ and $p(\tk,\ti/2)$;
~
this is used in lines 13.--15.\ and 21.--23., respectively.

Equations (\ref{pvHoare}) justify the steps in lines 13.--15.\ and
19.--21.;
equation (\ref{dataInv}) justifies step 7.--9.

\begin{enumerate}
\setlength{\itemsep}{-0.0cm}
\renewcommand{\labelenumi}{\arabic{enumi}.}
 \item{\2\verb||%
	$0 \leq {\tt x} < \SUM{0}$	
}\item{\1\verb|int find(int x) {				|
}\item{\1\verb|    int i, k, pv;				|
}\item{\2\verb|    |%
	$\SUM{N} \leq {\tt x} < \SUM{0}$	
}\item{\1\verb|    k  = 0;					|
}\item{\1\verb|    i  = N/2;					|
}\item{\2\verb|    |%
	$\SUM{\tk+2\cdot\ti} \leq {\tt x} < \SUM{\tk}
	\wedge p(\tk,\ti)
	\wedge \ti \geq 1$	
}\item{\1\verb|    pv = s[N/2];				|
}\item{\2\verb|    |%
	$\SUM{\tk+2\cdot\ti} \leq {\tt x} < \SUM{\tk}
	\wedge {\tt pv} = \SUM{\tk+\ti}
	\wedge p(\tk,\ti)
	\wedge \ti \geq 1$	
}\item{\1\verb|    while (i >= 2) {				|
}\item{\2\verb|        |%
	$\SUM{\tk+2\cdot\ti} \leq {\tt x} < \SUM{\tk}
	\wedge {\tt pv} = \SUM{\tk+\ti}
	\wedge p(\tk,\ti)
	\wedge \ti \geq 2$	
}\item{\1\verb|        if (x < pv) {				|
}\item{\2\verb|            |%
	$\SUM{\tk+2\cdot\ti} \leq {\tt x} < \SUM{\tk+\ti}
	\wedge {\tt pv} = \SUM{\tk+\ti}
	\wedge p(\tk,\ti)
	\wedge \ti \geq 2$	
}\item{\1\verb|            pv = pv + S(k+i*3/2) - s[k+i];	|
}\item{\2\verb|            |%
	$\SUM{\tk+2\cdot\ti} \leq {\tt x} < \SUM{\tk+\ti}
	\wedge {\tt pv} = \SUM{\tk+3\cdot\ti/2}
	\wedge p(\tk+\ti,\ti)
	\wedge \ti \geq 2$	
}\item{\1\verb|            k  = k + i;			|
}\item{\2\verb|            |%
	$\SUM{\tk+\ti} \leq {\tt x} < \SUM{\tk}
	\wedge {\tt pv} = \SUM{\tk+\ti/2}
	\wedge p(\tk,\ti)
	\wedge \ti \geq 2$	
}\item{\1\verb|        } else {				|
}\item{\2\verb|            |%
	$\SUM{\tk+\ti} \leq {\tt x} < \SUM{\tk}
	\wedge {\tt pv} = \SUM{\tk+\ti}
	\wedge p(\tk,\ti)
	\wedge \ti \geq 2$	
}\item{\1\verb|            pv = pv + S(k+i/2);		|
}\item{\2\verb|            |%
	$\SUM{\tk+\ti} \leq {\tt x} < \SUM{\tk}
	\wedge {\tt pv} = \SUM{\tk+\ti/2}
	\wedge p(\tk,\ti)
	\wedge \ti \geq 2$	
}\item{\1\verb|        }					|
}\item{\2\verb|        |%
	$\SUM{\tk+\ti} \leq {\tt x} < \SUM{\tk}
	\wedge {\tt pv} = \SUM{\tk+\ti/2}
	\wedge p(\tk,\ti/2)
	\wedge \ti \geq 2$	
}\item{\1\verb|        i = i/2;				|
}\item{\2\verb|        |%
	$\SUM{\tk+2\cdot\ti} \leq {\tt x} < \SUM{\tk}
	\wedge {\tt pv} = \SUM{\tk+\ti}
	\wedge p(\tk,\ti)
	\wedge \ti \geq 1$	
}\item{\1\verb|    }						|
}\item{\2\verb|    |%
	$\SUM{\tk+2} \leq {\tt x} < \SUM{\tk}
	\wedge {\tt pv} = \SUM{\tk+1}
	\wedge \ti = 1$	
}\item{\1\verb|    if (x < pv) {				|
}\item{\2\verb|        |%
	$\SUM{\tk+2} \leq {\tt x} < \SUM{\tk+1}
	\wedge {\tt pv} = \SUM{\tk+1}$	
}\item{\1\verb|        k  = k + 1;				|
}\item{\2\verb|        |%
	$\SUM{\tk+1} \leq {\tt x} < \SUM{\tk}$	
}\item{\1\verb|    } else {					|
}\item{\2\verb|        |%
	$\SUM{\tk+1} \leq {\tt x} < \SUM{\tk}$	
}\item{\1\verb|    }						|
}\item{\2\verb|    |%
	$\SUM{\tk+1} \leq {\tt x} < \SUM{\tk}$	
}\item{\1\verb|    return k;					|
}\item{\1\verb|}						|
}
\end{enumerate}

This completes the verification proofs of the algorithms given in
Fig.~\ref{Algorithms}.
\\
A short version of this paper (without proofs) was published in
\cite{Burghardt.2001}.

\bibliographystyle{alpha}
\bibliography{lit}


\end{document}